\documentclass[letterpaper, 10 pt, conference]{ieeeconf}  

\IEEEoverridecommandlockouts                              

\usepackage{amsmath} 
\usepackage{amssymb}  
\usepackage{bm}
\usepackage{graphicx}
\usepackage{multirow}
\usepackage{subfigure}
\usepackage{bm}
\usepackage{multicol}
\usepackage[T1]{fontenc}
\usepackage{cuted}
\usepackage[ngerman]{babel}
\usepackage[ansinew]{inputenc}
\usepackage[hyperindex=true, %
pdftitle={Gaussian Process-Based Model Predictive Control of Blood Glucose for Patients with Type 1 Diabetes Mellitus},%
pdfauthor={Lukas Ortmann, Dawei Shi, Eyal Dassau, Francis J. Doyle III, Steffen Leonhardt, Berno J.E. Misgeld},%
colorlinks=true,%
pagebackref=false,%
plainpages=false,%
pdfpagelabels,%
linkcolor=black,%
citecolor=black,%
filecolor=black,%
urlcolor=black%
]{hyperref} 
\usepackage[dvipsnames]{xcolor}
\usepackage[absolute,overlay]{textpos}
\usepackage{url}

\title{\LARGE \bf
Automated Insulin Delivery for Type 1 Diabetes Mellitus Patients using Gaussian Process-based Model Predictive Control
}

\author{Lukas Ortmann$^{1}$, Dawei Shi$^{2}$, Eyal Dassau$^{2}$, Francis J. Doyle III$^{2}$, Berno J.E. Misgeld$^{1}$, Steffen Leonhardt$^{1}$ 
	\thanks{This work was partially funded by the German Academic National Foundation and by the United States National Institutes of Health under grants DP3DK104057 and UC4DK108483.} 
	\thanks{$^{1}$Philips Chair for Medical Information Technology, 
		Helmholtz-Institute, RWTH Aachen University, 52074 Aachen, Germany.
		{\tt\small lukas.ortmann@rwth-aachen.de}}
	\thanks{$^{2}$Harvard John A. Paulson School of Engineering \& Applied Sciences, Harvard University, Cambridge, MA 02138, USA.}
}

\begin{document}

\maketitle
\thispagestyle{empty}
\pagestyle{empty}

\begin{textblock*}{\textwidth}(15mm,10mm) 
	\centering \bf \textcolor{NavyBlue}{Published on \emph{American Control Conference (ACC),} July 2019.\\\url{https://doi.org/10.23919/ACC.2019.8815258}}
\end{textblock*}

\begin{abstract}

The human insulin-glucose metabolism is a time-varying process, which is partly caused by the changing insulin sensitivity of the body. This insulin sensitivity follows a circadian rhythm and its effects should be anticipated by any automated insulin delivery system. This paper presents an extension of our previous work on automated insulin delivery by developing a controller suitable for humans with Type 1 Diabetes Mellitus. Furthermore, we enhance the controller with a new kernel function for the Gaussian Process and deal with noisy measurements, as well as, the noisy training data for the Gaussian Process, arising therefrom. This enables us to move the proposed control algorithm, a combination of Model Predictive Controller and a Gaussian Process, closer towards clinical application. Simulation results on the University of Virginia/Padova FDA-accepted metabolic simulator are presented for a meal schedule with random carbohydrate sizes and random times of carbohydrate uptake to show the performance of the proposed control scheme.

\end{abstract}
\smallbreak

\section{Introduction}
\label{sec:Introduction}

Approximately 415 million people had diabetes mellitus in 2015 and this number is assumed to increase to 642 million by the year 2040~\cite{Ogurtsova2017}. Of these patients, around 10\% have type 1 diabetes mellitus and they are therefore not able to control their blood glucose (BG) level without exogenous insulin injections. On the one side, the goal of these exogenous injections is to prevent hyperglycemia (BG~\textgreater~180~mg/dl) and secondary complications arising therefrom. On the other side, the induced insulin can lead to insulin-induced hypoglycemia (BG~\textless~70~mg/dl), which can be life-threatening. An automated insulin delivery device can help the patients achieve both these goals and patients do not need to control there BG level manually anymore. Recent developments in insulin pumps and glucose sensors enable us to solve the problem of insulin delivery in closed-loop, which leads to improved BG regulation and which is the goal of the artificial pancreas project~\cite{Thabit2016},~\cite{Doyle2014}.

A variety of factors in the insulin-glucose metabolism change over time and call for adaptive control techniques. These factors are e.g. the irregular pattern of exercises~\cite{Zinman2003}, the quasi-periodic appearance of meals~\cite{Wang2010} and the diurnal changes  of the insulin sensitivity~\cite{Toffanin2013}.
There are many proposed control algorithms in the literature that focus on the requirement of adaptation. Based on the patient's reaction to boluses given during the last days, the carbohydrate (CHO) to insulin ratio is adapted in~\cite{Palerm2007}, to enhance the rejection of meal intakes.
Run-to-run control was also used in~\cite{Palerm2008} to improve the tracking performance and to control the blood glucose level to a tighter zone, by updating the basal infusion rate of patients based on past measurements.
In comparison to these studies, where only a few BG measurements were available during the day, the following studies used continuous glucose measurements. The periodic appearance of meals was used in~\cite{Wang2010} to improve the control performance with a Model Predictive Iterative Learning Controller.
In~\cite{Toffanin2017}, run-to-run adaptation of the basal rate was used during night time, while the carbohydrate-to-insulin ratio, which is used to calculate meal boluses, was adapted during the day.
The control penalty in the cost function of a zone MPC was adapted in~\cite{Shi2018} to reduce the mean glucose level, while not increasing the risk of hypoglycemia.
Both insulin and the counterregulatory hormone glucagon were used in a clinical study on pigs, where the authors used a Generalized Predictive Control approach~\cite{El-Khatib2007}.
In~\cite{Colmegna2016}, the controller was adapted by Gain Scheduling based on the blood glucose concentration~\cite{Colmegna2016}, whereas the controller was switched based on an estimate of the insulin sensitivity in~\cite{Misgeld2016b}.
Another publication where the insulin sensitivity is included into the controller to enhance the control performance is~\cite{Toffanin2013}, where the insulin sensitivity was included into the input constraint of a Model Predictive Controller (MPC).
In~\cite{Toffanin2018} the changing insulin sensitivity was included into a run-to-run controller, which adjusts the carbohydrate-to-insulin ratio and the basal insulin delivery rate.
The authors in~\cite{Turksoy2014} provide a review of adaptive controllers for BG regulation, including Self-Tuning-Regulators, Minimum Variance Control, Generalized Predictive Control and Linear Quadratic Regulators.
Our idea of facilitating Gaussian Processes in the field of glucose control is also used in \cite{Messori2016} to determine personalized linear patient models.

\begin{figure}
	\centering
	{\includegraphics*[width=\columnwidth,trim={5cm 10cm 12.5cm 2.75cm},clip]{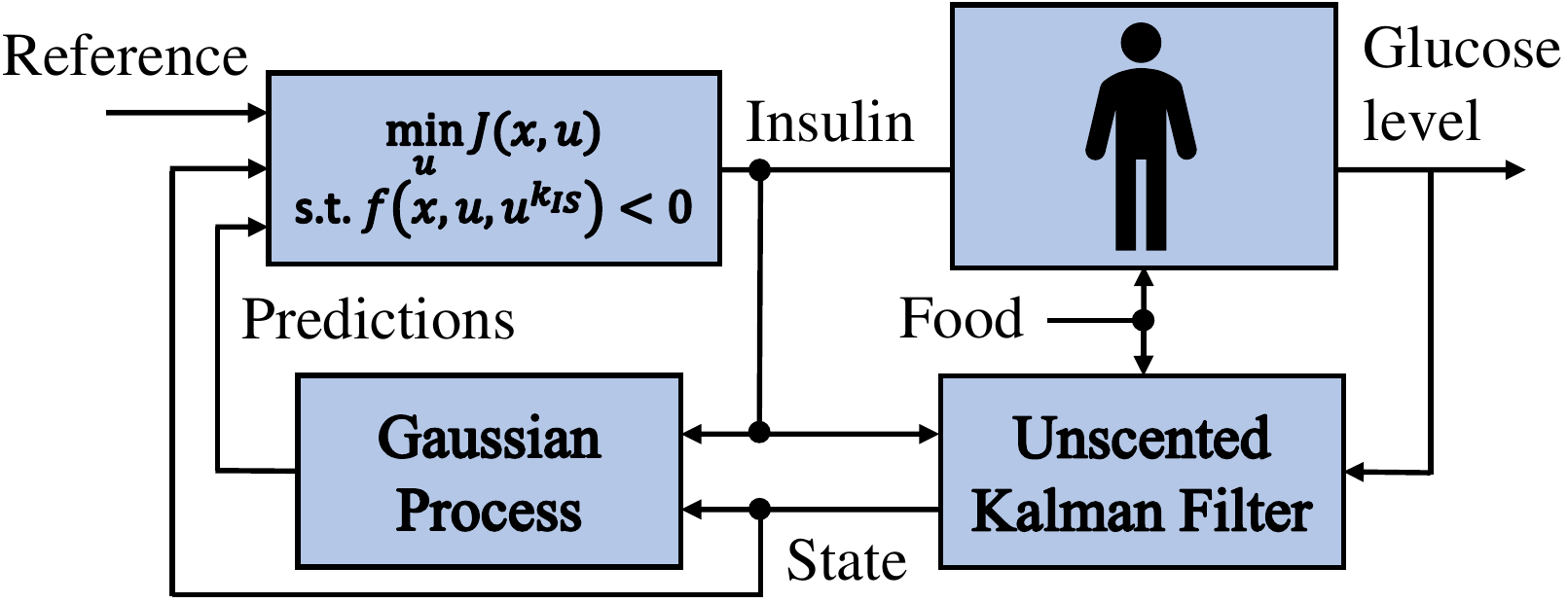}}
	\caption{Block diagram of the proposed controller, consisting of an Unscented Kalman Filter, the Gaussian Process and a Model Predictive Controller.}
	\label{fig:blockdiagram}
\end{figure}

In this paper, we present an extension of our work on incorporating information about the changing insulin sensitivity into a controller for the insulin-glucose metabolism \cite{Ortmann2017}. In contrast to previously published results on including the insulin sensitivity, we determine the effect of the changing insulin sensitivity during closed-loop. Due to the periodicity of the insulin sensitivity, we can anticipate the upcoming effect of the changing insulin sensitivity by using the collected data. We extract the effect of the changing insulin sensitivity from the state estimate which is provided by an Unscented Kalman Filter \cite{Misgeld2017}. The collected data of the effect of the changing insulin sensitivity is given to a Gaussian Process which predicts the future values of the effect. These predictions are then given to a MPC which incorporates the information into the optimization problem, when calculating the insulin injections. A block diagram of the proposed control structure is provided in Fig.~\ref{fig:blockdiagram}.

The newly developed content of this paper is that we present a controller parametrized for humans, in comparison to the original controller, which was using a model for G{\"o}ttingen Minipigs. We also provide a new kernel function for the Gaussian Process that is fading out old data and is capable of dealing with the measurement noise we are facing in applications and our simulation model. By postprocessing the collected training data for the Gaussian Process, the controller becomes insensitive to unannounced meals and meal sizes and times can be arbitrary. To show the advantages of including a learning part into the glucose controller we show simulation results on the FDA-accepted University of Virginia/Padova (UVA/Padova) metabolic simulator~\cite{Dalla2014}.

The paper is structured as follows: In Section~\ref{sec:review} we give a review of our previous work. We explain how we adapt the metabolic simulator to show a changing insulin sensitivity and present the model used inside the MPC in~\ref{sec:system_modeling}. The postprocessing of the training data, the new kernel function for the Gaussian Process and its predictions are shown in Section~\ref{sec:Gaussian_Process}. Afterwards, the results are shown in Section~\ref{sec:Results} and we give a conclusion in Section~\ref{sec:Conclusion}.

\section{Review of Previous Work}
\label{sec:review}
Here, we give an overview of our previous work. We explain how the training data is extracted from the state estimates and the insulin injections. Furthermore, we describe the Model Predictive Controller that is provided with the predictions of the Gaussian Process.

\subsection{Training Data Calculation}
\label{subsec:training_data_collection}

The training data for the Gaussian Process is calculated during closed-loop control from the state estimate of the UKF and the insulin input. Every new glucose measurement gives a new state estimate, which is then used to derive a new training data point. To do so, we start with the time-varying linear model given in \eqref{eq:model}, that we split up into a model describing the insulin glucose metabolism ($\bm{\hat{A}}\bm{x}(t)$) and a part that describes the effect of the changing insulin sensitivity ($\bm{A}^{k_{\textrm{IS}}}(t)\bm{x}(t)$):
\begin{equation}
\begin{split}
\bm{\dot{x}}(t)	&=\bm{A}(k_{\textrm{IS}}(t))\bm{x}(t)+\bm{B}u(t)\\
&=\bm{\hat{A}}\bm{x}(t)+\bm{A}^{k_{\textrm{IS}}}(t)\bm{x}(t)+\bm{B}u(t).
\end{split}
\end{equation}
The second part is then interpreted as a disturbance $u^{k_{\textrm{IS}}}(t)$ induced by the changing insulin sensitivity, which enters the system through $\bm{B}^{k_{\textrm{IS}}}$. This leads to:
\begin{equation}\label{eq:linear_system}
\bm{\dot{x}}(t)=\bm{\hat{A}}\bm{x}(t)+\bm{B}u(t)+\bm{B}^{k_{\textrm{IS}}}u^{k_{\textrm{IS}}}(t).
\end{equation}
This system is now discretized and the notation $[\cdot]_i$ is introduced to refer to the $i$th row of a vector/matrix variable. Furthermore, we denote $[\bm{x}]_{i^*}$ as the row in the A-matrix of \eqref{eq:model} which is time-varying. With the discretized system,
\begin{equation}\label{Eq:Linear_Model}
\bm{x}_{k+1}=\bm{\hat{A}}_{d}\bm{x}_{k}+\bm{B}_{d}u_{k}+\bm{B}^{k_{\textrm{IS}}}_{d}u^{k_{\textrm{IS}}}_{k},
\end{equation}
we calculated the disturbance as follows:
\begin{equation}
u^{k_{\textrm{IS}}}_{k-1}=\left(\left[\bm{x}_{k}\right]_{i^*}-\left[\bm{\hat{A}}_{d}\bm{x}_{k-1}\right]_{i^*}-\left[\bm{B}_{d}u_{k-1}\right]_{i^*}\right)/\left[\bm{B}^{k_{\textrm{IS}}}_{d}\right]_{i^*}.
\end{equation}
The training data for the Gaussian Process consists of these data points and there corresponding time stamps. For more details, see~\cite{Ortmann2017}.

\subsection{Controller}
\label{subsec:Controller}

The controller that we are using is a MPC that is supplied with the information about the predicted effect of the changing insulin sensitivity $u_k^{k_{\textrm{IS}}}$. We will denote this combination of Gaussian Process and MPC as GP-MPC. The Gaussian Process is predicting the effect for the complete prediction horizon of the MPC formulation. If we are referring to MPC, we mean the same controller but without the predictions of the Gaussian Process.
The GP-MPC is defined as follows:
\begin{align}\label{eq:MPC}
\begin{split}
J^*_{0\rightarrow N-1}(&\bm{x}_0)=\min_{\bm{U}_{0\rightarrow N-1}} \;J_{0\rightarrow N-1}(\bm{x_0},\bm{U}_{0\rightarrow N-1})\\
\text{s.t.} \; \bm{x}_{k+1}&=\bm{\hat{A}}_{d}\bm{x}_k+\bm{B}_{d}u_k+\bm{B}^{k_{\textrm{IS}}}_{d}u_k^{k_{\textrm{IS}}},\\
\bm{x}_0&=\bm{x}(t),\quad y_k=\bm{C}_{d}\bm{x}_k,\quad y_N= 0,\\
0 &\leq u_k+u_\textrm{basal}\leq u_\textrm{max},
\end{split}
\end{align}
where the cost function is
\begin{equation}
J_{0\rightarrow N-1}(\cdot,\cdot)={\sum_{k=0}^{N-1}} y^T_{k}Qy_{k}+(u_{k}-u^{ss}_k)^TR(u_{k}-u^{ss}_k).
\end{equation}

The matrices can be obtained by discretizing the system in \eqref{eq:model} for the nominal insulin sensitivity ($k_\textnormal{IS}(t)=1$) and they describe the linearization of the model around the steady state $\bm{x}_\textnormal{basal}$ for the basal input $u_\textnormal{basal}$. Therefore, when the MPC steers the state $\bm{x}_k$ to zero, this corresponds to the basal conditions of the patient. The vector $\bm{x}_k$ is the derivation of the state from the linearization point $\bm{x}_\textnormal{basal}$, $u_k$ is the derivation of the insulin injection form its basal rate $u_\textnormal{basal}$ in mU/min, $u_k^{k_{\textrm{IS}}}$ is the predicted influence of the changing insulin sensitivity provided by the Gaussian Process, $y_N$ is the blood glucose level in mg/dl, $N$ is the prediction horizon of the MPC and $Q$ and $R$ are the weights of the MPC.
To enable reference tracking in the presence of the disturbance $u^{k_{\textrm{IS}}}$, we introduce $u^{ss}_k$, that is needed to reject the disturbance $u^{k_{\textrm{IS}}}_k$ at steady state and is calculated by solving the linear equations
\begin{equation}
\begin{bmatrix}
\bm{\hat{A}}_d-\bm{I} & \bm{B}_d \\
\bm{C}_d & 0
\end{bmatrix}
\begin{bmatrix}
\bm{x}^{ss}_k \\
u^{ss}_k
\end{bmatrix} =	
\begin{bmatrix}
-\bm{B}^{k_{\textrm{IS}}}_du^{k_{\textrm{IS}}}_k \\
0
\end{bmatrix},
\end{equation}
$\forall k \in \left[0,N-1\right]$.
The constraint finite time optimal control (CFTOC) problem in \eqref{eq:MPC} is solved with the Yalmip toolbox~\cite{Lofberg2004} and the first element of the solution is given as a command to the insulin pump and the optimization problem is solved again once a new measurement is available. The parameters of the MPC are chosen as follows. The prediction horizon is $N=30$ and the sample time of the model in the MPC is 5~minutes. The prediction horizon is therefore 2.5 hours, which covers the most important dynamics of the insulin-glucose metabolism. The weights of the MPC are chosen as $Q=1$ and $R=10$ and the input is constraint to a maximum of $u_\textrm{max}=0.1$~U/min.

\section{System Modeling}
\label{sec:system_modeling}
In this section, we describe how we adapt the simulation model which we use to evaluate the control performance, to include a changing insulin sensitivity. We also describe the changes made to the model within the MPC, such that it describes the human insulin-glucose metabolism.
\subsection{Simulation Model}
\label{subsec:Simulation_Model}
To evaluate the performance of the proposed control algorithm we use the UVA/Padova simulator. The constant insulin sensitivity of the simulator is adapted such that the patient shows a diurnal change in insulin sensitivity.
To do so, we follow the approach in~\cite{Visentin2015} and~\cite{Toffanin2018}, where the parameters $V_{mx}$ and $k_{p3}$ are changed depending on the insulin sensitivity:
\begin{equation}
\begin{split}
V_{mx}(t)&=V_{mx}^\textnormal{nominal}\cdot k_{\textnormal{IS}}(t)\\
k_{p3}(t)&=k_{p3}^\textnormal{nominal}\cdot k_{\textnormal{IS}}(t).
\end{split}
\end{equation}
We use the insulin sensitivity curve in Fig.~\ref{fig:training_data_prediction_and_k_IS}, which is based on data from~\cite{Toffanin2013}.

\subsection{Controller Model}
\label{subsec:Controller_Model}

The model used inside the Model Predictive Controller is based on the Lunze model~\cite{Lunze2014b}. This model is a simplification of the Sorensen model~\cite{Sorensen1985} and was originally developed for G{\"o}ttingen Minipigs. The advantage of this model, for our purposes, is that the insulin sensitivity is specifically modeled by its own parameter and only effects one state of the model. This makes it easier to extract the effect of the changing insulin sensitivity from the state estimate. A reduced order nonlinear version of the Lunze model \cite{Misgeld2016a} is used in the Unscented Kalman Filter and a linearized version is used inside the CFTOC problem of the MPC and to obtain the training data for the Gaussian Process. The Lunze model was completely reparametrized to adapt it to the insulin-glucose metabolism of humans by using input-output data of the Dalla Man model and publicly available data from~\cite{Kovatchev2010}. We choose to adapt the Lunze model to the Dalla Man model, because it is accepted and has been used to support FDA regulated clinical studies. The gastro-intestinal tract and the subcutaneous insulin route were reparametrized using the data from~\cite{Kovatchev2010}, while the remaining parameters were obtained using input-output data of patient~1 of the simulator.

\section{Gaussian Process}
\label{sec:Gaussian_Process}

This section presents the newly developed postprocessing of the training data and the new combination of kernel functions. We also show the collected training data and the predictions made by the newly parametrized Gaussian Process.

\subsection{Postprocessing the Training Data}
\label{subsec:postprocessing}
Food intakes disturb the training data calculation and should therefore be announced to the controller. Due to model mismatch in the gastro-intestinal tract, the error due to the food intake cannot be completely calculated and we therefore discard training data that are collected after an announced food intake. Using the model of the gastro-intestinal tract, we disable the training data collection, while the rate of glucose entering the blood stream is larger than 150~mg/min.

Furthermore, we discard training data points which are out of range, which we define by $|u^{k_{\textrm{IS}}}_{k}|>2$~mg/dl. This second trigger for discarding training data is important to cope with forgotten meal announcements, so called unannounced meals. Otherwise, the unannounced rise in the glucose level will be interpreted as a drop in insulin sensitivity (high $u^{k_{\textrm{IS}}}_{k}$ values) and therefore corrupts the training data.

It is also possible to discard data points when the patient announces an exercise or unusual physical activity, because these drain glucose from the system and would alter the training data. Also, more advanced techniques could be used to determine if the collected training data points should be discarded and it is also possible to let the patient interact with the controller to benefit form the patients knowledge and experience. For example, if there is a large peak in the training data values, the controller could ask the patient if he/she forgot to announce a meal and if so what the meal size was. With the information about the meal size, the corrupted training data can then be recalculated.

\subsection{Kernel Function and Predictions}
\label{subsec:kernel_and_predictions}
Once the training data is collected, it is used by the Gaussian Process to predict future values of the effect of the changing insulin sensitivity. These values can then be included in the CFTOC problem of the MPC. We use the following two kernel functions to define the correlation between our data points. Namely, a periodic kernel function and an exponential kernel function:
\begin{equation}
k_\textrm{{P}}(t,t';l_\textrm{{P}},\lambda)=\exp\left((-2\sin^2\left(\dfrac{\pi}{\lambda}(t-t')\right))/l_\textrm{{P}}^2\right).
\end{equation}
\begin{equation}
k_\textrm{{E}}(t,t';l_\textrm{{E}})=\exp\left((-|t-t'|)/l_{\textrm{E}}\right)
\end{equation}
The periodic kernel functions enables us to have a high correlation between data points that were collected at the same time of day. The exponential kernel function gives us the opportunity to fade out old data points over time by reducing their correlation to the prediction. Finally, we add a noise term to the kernel function, because the calculated training data is affected by the measurement noise. This leads to:
\begin{equation}
k_\textrm{C}(t,t';\bm{\eta})=\theta^2\cdot k_\textrm{{E}}(t,t';l_\textrm{E}) \cdot k_\textrm{P}(t,t';l_\textrm{P},\lambda)+\sigma_n^2.
\end{equation}
The hyperparameters of the kernel function are lumped into $\bm{\eta}:=[\theta^2,l_\textrm{{E}},l_\textrm{{P}},\lambda,\sigma_n]$ and their values can be seen in Table~\ref{tab:hyperparameters}. The periodic length, $\lambda$, is chosen to be equal to the length of the diurnal insulin sensitivity rhythm, which is 24~hours. The length scale of the exponential kernel function, $l_\textrm{{E}}$, is chosen such that the correlation of data points that are 3~days old is reduced to 90\%. The length scale of the periodic kernel function, $l_\textrm{{P}}$, and the variance, $\theta$, are determined through hyperparameter optimization and then fixed at these values. Finally, the noise standard deviation, $\sigma_n$ is adapted to the noise level.

\begin{table}[]
	\centering
	\caption{Hyperparameters of the Gaussian Process.}
	\begin{tabular}{c|ccc}
		\hline
		\hline
		Parameter 	& $\theta$ [1]	& $\lambda$ [min]	& $l_\textnormal{P}$ [1]\\
		\hline
		Value 		& $0.071$		& $1440$			& $0.549$\\
		\hline
		Parameter 	& $l_\textnormal{E}$ [min]	& $\sigma_n$ [1] 	&\\
		\hline
		Value 		& $4.1\cdot10^4$			& $0.2$ 			&\\
		\hline
		\hline
	\end{tabular}
	\label{tab:hyperparameters}
\end{table}

\begin{figure}
	\centering
	{\includegraphics*[width=\columnwidth,trim={4cm 8.5cm 3.55cm 8.5cm},clip]{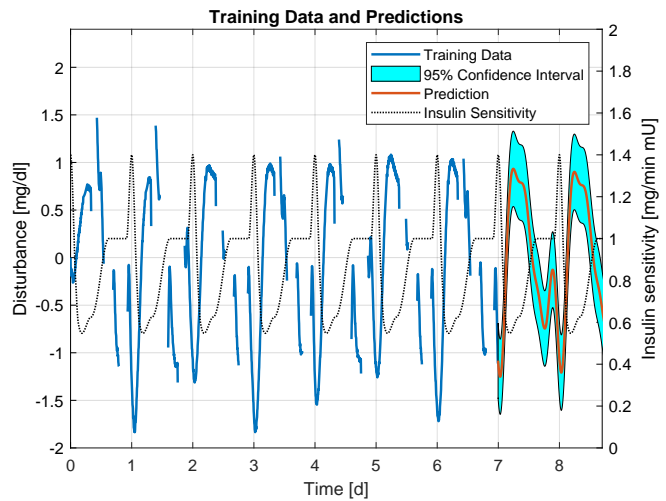}}
	\caption{Training data for the Gaussian Process and its prediction and variance, as well as, the insulin sensitivity rhythm. The training data is disjoint, because of the postprocessing explained in Section~\ref{subsec:postprocessing}.}
	\label{fig:training_data_prediction_and_k_IS}
\end{figure}

A set of training data points is presented in Fig.~\ref{fig:training_data_prediction_and_k_IS}. In the figure, it can be seen that training data points are discarded (either because a meal input is present or because their absolute value is out of range). Furthermore, we can see the prediction of the Gaussian Process using this data and the variance of the prediction. Please note that only the predictions for the next 2.5~hours are going to be used in the CFTOC problem and that the peaks in the training data are not present in the prediction, because these peaks are averaged out.

\begin{figure*}
	
	\arraycolsep=1.8pt
	\begin{equation*}
	A(k_\textrm{IS}(t))=\left[\begin{array}{cccccccccccc}
	-0.70  &  0.32  &  0.38    &     0      &   0    &     0    &     0    &     0   &      0    &     0  &  0.024  &  0\\[-.075cm]
	0.50 &  -0.56   &      0     &    0   &-0.010 &   5.11  & -5.63  &  5.62  &       0  & -0.030     &    0   &  0\\[-.075cm]
	1.45      &   0  & -2.75 &   1.30  &       0   &      0  &       0   &      0   &      0       &  0       &  0  &   0\\[-.075cm]
	0    &     0  &  0.20  & -0.20&   -0.025 \cdot k_\textrm{IS}(t) &       0   &      0   &      0    &     0  &       0 &        0  &    0\\[-.075cm]
	0    &     0   &      0     &    0  & -0.091    &     0   &      0     &    0    &     0  &  0.075    &     0   &   0\\[-.075cm]
	-0.0009    &     0     &    0     &    0 &  -0.0005  & -0.08    &     0     &    0  &       0  & -0.0015    &     0   &   0\\[-.075cm]
	0     &    0  &       0   &      0    &     0 &   0.007 &  -0.015   &      0      &   0   &      0      &   0   &  0\\[-.075cm]
	0      &   0   &      0     &    0 &  -0.0009   &      0     &    0 &  -0.04   &      0  & -0.0028   &      0  &   0\\[-.075cm]
	0       &  0     &    0       &  0       &  0       &  0        & 0      &   0 &  -0.025  &       0    &     0   &  0\\[-.075cm]
	0       &  0        & 0      &   0    &     0        & 0        & 0      &   0   & 0.011 &  -0.011      &   0  &   0\\[-.075cm]
	0        & 0        & 0         &0      &   0     &    0        & 0         &0     &    0     &    0   &-0.078   &  0.0078\\[-.075cm]
	0       &  0      &   0   &      0       &  0     &    0       &  0     &    0       &  0   &      0      &   0   &  -0.0077\\[-.075cm]
	\end{array}\right],
	\end{equation*}
	\arraycolsep=4pt
	\begin{align}
	\label{eq:model}
	\quad B = \left[\begin{array}{cccccccccccc} 0& 0& 0& 0& 0& 0& 0& 0& 0.0216& 0.0014& 0& 0 \end{array}\right]^T,
	\quad
	C = \left[\begin{array}{cccccccccccc} 1 & 0 & 0 & 0 & 0 & 0 & 0 & 0 & 0 & 0 & 0 & 0 \end{array}\right]
	\end{align}

\end{figure*}

\section{Results}
\label{sec:Results}

\setlength\tabcolsep{3pt}
\begin{table}
	\centering
	\caption{Results with announced meals.}
	\begin{tabular}{l|cc|cc}
\hline \hline
&\multicolumn{2}{c|}{Day and night} &\multicolumn{2}{c}{Overnight} \\
Metric & MPC & GP-MPC & MPC & GP-MPC \\
\hline\% time & & & &\\
\quad<54 mg/dl & 0.0& 0.0& 0.0& 0.0 \\
\quad<60 mg /dl &  0.3& 0.0& 1.1& 0.0  \\
\quad<70 mg/dl &  2.4& 0.3&7.8& 1.0\\
\quad70-140 mg/dl &  45.6& 47.4& 79.0& 76.1  \\
\quad70-180 mg/dl &  72.1& 75.2& 91.5& 98.8 \\
\quad>180 mg/dl &  24.8& 24.0& 0.0& 0.0\\
\quad>250 mg/dl &  0.2& 1.0& 0.0& 0.0\\
\quad>300 mg/dl &  0.0& 0.0& 0.0& 0.0\\
Mean glucose [mg/dl] &  146.4& 149.2&109.5& 120.6\\
Median glucose [mg/dl] &  141.0& 141.0& 109.0& 123.0 \\
SD glucose [mg/dl] &  44.9& 40.9& 26.3& 22.3\\
Coefficient of glucose variation & 0.3& 0.3& 0.2& 0.2\\
Mean glucose at 7:00am [mg/dl] &  134.1& 125.3& ---&--- \\ 
\hline \hline
\end{tabular}
\label{tab:results}
\end{table}

To evaluate the performance of the proposed control scheme, we use the UVA/Padova simulator and compare the developed GP-MPC with the MPC (same controller but without the predictions of the Gaussian Process). We simulate meal intakes with randomized timing and carbohydrate sizes for patient 1 of the simulator. This patient has a fasting glucose level of 122~mg/dl, which is the control reference, and a basal insulin injection of 20.4~mU/min. To keep the meal times and sizes reasonable, breakfast is eaten between 7am and 9am and has sizes between 40~g and 60~g. Lunch is eaten between 12pm and 2pm and dinner is eaten between 6pm and 8pm, both with a size between 60~g and 90~g. The exact meal sizes and meal times for the 7~day simulation can be found in Table~\ref{tab:meals}. We first present the performance for announced meals and then show how the closed-loop system is behaving when a patient forgets to announce a meal.

\setlength\tabcolsep{1.7pt}

\begin{table}[]
	\centering
	\caption{Meal sizes in \textnormal{g} carbohydrates and meal times.}
	\begin{tabular}{c|ccccccc}
		\hline
		\hline
		& Day 1 & Day 2 & Day 3 & Day 4 & Day 5 & Day 6 & Day 7\\
		\hline
		\multirow{2}{*}{Breakfast} 	& 50 g & 40 g	&60 g	&55 g 	& 50 g	&40 g &60 g\\
									& 8am  & 7:30am &9am 	&8:30am & 9am 	& 7am & 7:30am\\
		\hline	
		\multirow{2}{*}{Lunch} 		& 90 g	&70 g	&80 g 		&75 g	&80 g	&70 g 	&85 g\\
									& 1pm 	& 12pm	&12:30pm	&1pm	&12pm	& 1:30pm	&12:30pm\\
		\hline
		\multirow{2}{*}{Dinner} 	& 75 g	&60 g 		&85 g  		&90 g 	&85 g  		&90 g 		&70 g\\
									& 7pm	& 7:30pm 	&6:30pm 	& 6pm	& 7:30pm	& 6:30pm	&8pm\\
		\hline
		\hline
	\end{tabular}
	\label{tab:meals}
\end{table}

\begin{figure}
	\centering
	{\includegraphics*[width=\columnwidth,trim={4.15cm 9.9cm 4.7cm 8.85cm},clip]{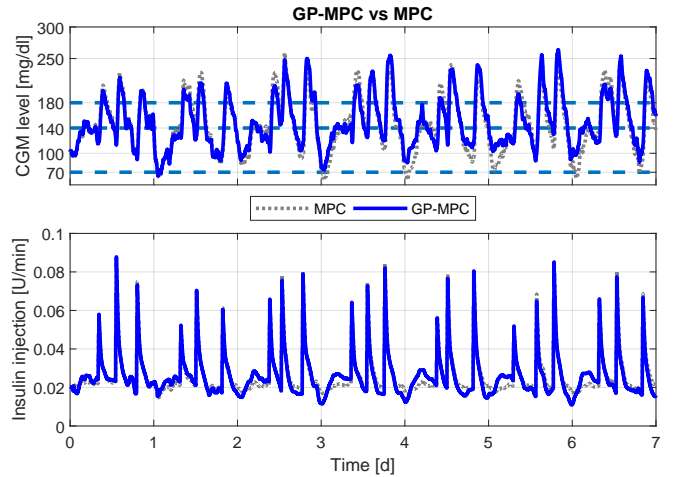}}
	\caption{Performance of GP-MPC and MPC for announced meals.}
	\label{fig:announced}
\end{figure}

\begin{figure}
\centering
{\includegraphics*[width=\columnwidth,trim={4.15cm 9.9cm 4.7cm 8.85cm},clip]{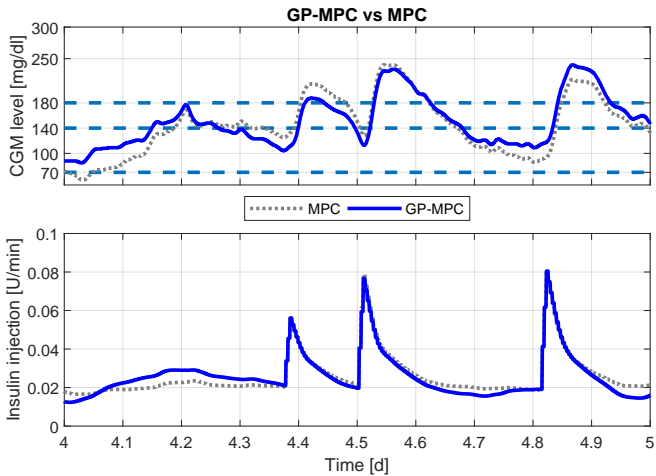}}
\caption{Closeup of day 5 for GP-MPC and MPC for announced meals.}
\label{fig:announced_closeup}
\end{figure}

The performance of the GP-MPC and MPC for announced meals can be seen in Fig.~\ref{fig:announced} and Fig.~\ref{fig:announced_closeup}. The upper panel of Fig.~\ref{fig:announced} shows that with the GP-MPC there are no events of hypoglycemia after the second night and the drops that can be seen for the MPC during the night are prevented. Furthermore, the glucose level around breakfast is lower and closer to the reference, which leads to lower glucose peaks after the patient eats breakfast. In the closeup in Fig.~\ref{fig:announced_closeup}, the performance improvement, coming from the predictions of the Gaussian Process, can be seen more clearly. The glucose level is closer to the reference value for most of the day. Only after dinner the glucose level rises higher than the one with the MPC, but this is due to the low glucose level with the MPC before dinner, which is 35~mg/dl below the fasting glucose level. The metrics in Table~\ref{tab:results} underline the performance advantage of the GP-MPC. There are less events of hypoglycemia and the standard deviation of the glucose level can be reduced. The time in range for 70-180~mg/dl can be increased during day and night through incorporating the Gaussian Process. For the range 70-140~mg/dl, the GP-MPC outperforms the MPC during the day. During the night, the value for the GP-MPC is slightly worse than for the MPC, which originates from the glucose level being closer to the fasting glucose level before dinner. The glucose level with the GP-MPC is higher before dinner, which leads to glucose level above 180~mg/dl during the night. In general, the set point for the GP-MPC could be lowered in comparison to the MPC, which would reduce the events of hyperglycemia, while not introducing events of hypoglycemia. This is another advantage of adopting insulin sensitivity anticipation.

\begin{figure}
	\centering
	{\includegraphics*[width=\columnwidth,trim={4.15cm 9.9cm 4.7cm 8.85cm},clip]{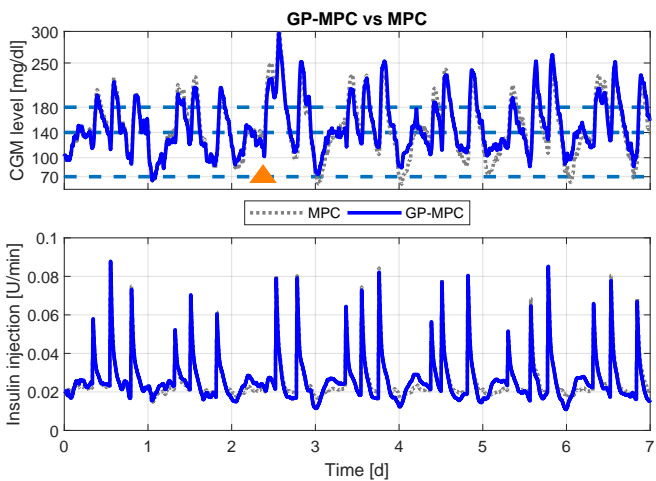}}
	\caption{Performance of GP-MPC and MPC for an unannounced breakfast~(60 g) on day 3, which is indicated by the orange triangle.}
	\label{fig:unannounced}
\end{figure}

\begin{figure}
	\centering
	{\includegraphics*[width=\columnwidth,trim={4.15cm 9.9cm 4.7cm 8.85cm},clip]{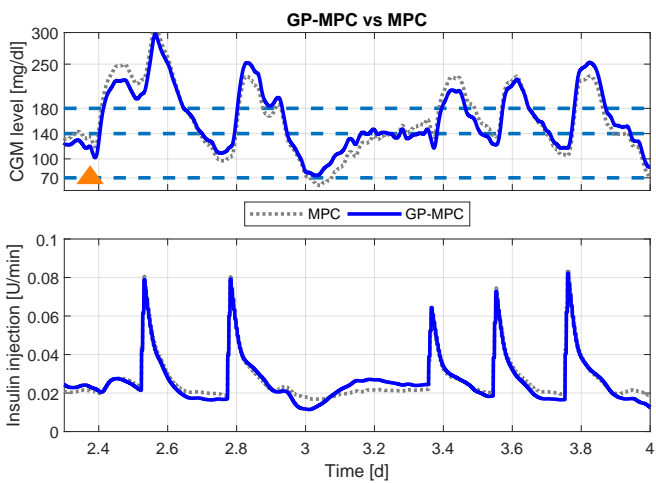}}
	\caption{Closeup for GP-MPC and MPC of the period after the unannounced meal, which is indicated by the orange triangle.}
	\label{fig:unannounced_closeup}
\end{figure}

In Fig.~\ref{fig:unannounced} and Fig.~\ref{fig:unannounced_closeup} we show how the controller is reacting to an unannounced meal of 60~g of CHO, which is indicated by the orange triangle. In the lower panels, one can see that the insulin bolus is missing, because the meal is not announced and in the upper panels one can see the subsequent large peak in the glucose level. This unannounced food intake leads to training data that is out of range and gets discarded, as explained in Section~\ref{subsec:postprocessing}. Therefore, the training data does not become corrupted and the controller will not introduce a peak in the insulin injections on the next day. In the closeup in Fig.~\ref{fig:unannounced_closeup}, the behavior of the controller during the following day can be seen and it shows that the control performance is not altered by the unannounced meal.



\section{Conclusion}
\label{sec:Conclusion}

This paper presents an enhancement to our previous work by developing a controller suitable for humans that is able to deal with measurement noise and unannounced meals. Now, the Gaussian Process fades out old training data and is insensitive to noise. Furthermore, the performance advantage of anticipating the changing insulin sensitivity is shown on a FDA-accepted simulation model, which leads the developed controller towards clinical application. The proposed integration of a controller with a machine learning technique has proven to be effective and could enhance other glucose control strategies. Finally, through learning the patient's specific insulin sensitivity circadian rhythm with fading memories, the proposed controller enables and improves personalized health care.





\bibliographystyle{IEEEtran}
\bibliography{IEEEabrv,ACC2019}

\end{document}